\documentclass[a4paper,11pt]{article}
\pdfoutput=1
\usepackage{latexsym}
\usepackage{amsmath}
\usepackage{amsfonts}
\usepackage{graphicx}

\def\laq{~\raise 0.4ex\hbox{$<$}\kern -0.8em\lower 0.62
ex\hbox{$\sim$}~}
\def\gaq{~\raise 0.4ex\hbox{$>$}\kern -0.7em\lower 0.62
ex\hbox{$\sim$}~}

\def\beq{\begin{equation}}
\def\eeq{\end{equation}}
\def\bea{\begin{eqnarray}}
\def\eea{\end{eqnarray}}
\def\bean{\begin{eqnarray*}}
\def\eean{\end{eqnarray*}}

\def \hp {\dot{h}}

\def\l {\langle}
\def\re {\rangle}
\def\eff{e\!f\!f}

\def \pa {\partial}
\def \ra {\rightarrow}
\def \ti {\widetilde}

\def \b {\beta}
\def \a {\alpha}

\def \ep {\epsilon}

   \def\be{\begin{equation}}
   \def\ee{\end{equation}}
   \def\ba{\begin{eqnarray}}
   \def\ea{\end{eqnarray}}

   \def\alphat{\tilde{\alpha}}
   \def\betat{\tilde{\beta}}
   
   \def\hp{\hat{\psi}}
   \def\vf{\varphi}

\title{Isotropic Observers and the Inflationary Backreaction Problem}

\author{Giovanni Marozzi$^1$\thanks{giovanni.marozzi@college-de-france.fr} \ and 
             Gian Paolo Vacca$^2$\thanks{vacca@bo.infn.it}\\
      \\ 
      ${}^1$  Coll\`ege de France, 11 Place M. Berthelot, 75005 Paris, France\\
      ${}^2$ INFN Sezione di Bologna, 
      via Irnerio 46, I-40126 Bologna, Italy
}
\date{}
\begin{document}
\addtolength{\belowdisplayskip}{-0.0cm}       
\addtolength{\abovedisplayskip}{-0.0cm}       

\maketitle
\begin{abstract} 
In an inflationary regime driven by a free massive inflaton we consider a class of observers which sees an inhomogeneous and isotropic Universe
and derive within a genuinely gauge invariant approach the backreaction effects due to long wavelength scalar fluctuations on the associated effective Hubble factor and equation of state. We find that, for such so-called isotropic observers, contrary to what happens for the
observables defined by free-falling observers, there is an effect to leading order in the slow-roll parameter  in the direction of slowing down the measured rate of expansion and of having an effective equation of state less de Sitter like. 
From a general point of view the isotropic observers result has to
be considered complementary to other cases (observers)
in helping to characterize the physical properties of the models under
investigation.
\end{abstract}

\section{Introduction}

The difficult task of probing quantum effects in the dynamics of our Universe is usually casted perturbatively as a backreaction problem.
The computation of backreaction effects 
both in the present (see \cite{ReviewsBE} for recent reviews) and in the early Universe is a 
long-standing problem related both to the evaluation and interpretation of the results and to 
the fundamental ambiguities
in constructing perturbatively gauge invariant (GI) 
observables ~\cite{Uc} and average quantities.

Following the basic fact that the averaging procedure does not commute with the nonlinear
evolution of Einstein equations~\cite{GE}, the dynamics of the averaged geometry turns out to be affected by so-called "backreaction" terms originating from the space-time inhomogeneities.
Such point was first exploited to study the
effective dynamics of the averaged geometry for a dust Universe~\cite{Buchert}, 
and recently used to evaluate, for the first time, in a GI way the quantum backreaction in a 
model of chaotic inflation \cite{FMVVprl}.
In fact, the computation of backreaction effects
induced by cosmological fluctuations
in an inflationary era ~\cite{ABM}, has been the subject of controversial analysis
\cite{Uc, ABM, All,FMVV_II} related to the gauge choice.
Such gauge problem has recently been addressed by a suitable GI but observer dependent averaging prescription \cite{GMV1} and by a general-covariant and GI formulation of the effective Einstein
equations for the cosmological backreaction \cite{GMV2}, which were the base of \cite{FMVVprl}.

Let us stress what we mean for {\it gauge invariance} in the
method used here to describe  the cosmological backreaction.
The goal of such a method is to give statements on the observables which are
constructed by averaging procedures with respect to a particular class
of observers.
Any physical class of observers is associated to a reference frame, therefore
from the point of the diffeomorphism gauge freedom a gauge fixing is intrinsically
necessary at physical level whenever an observer is considered. This is
simply unavoidable, since we know that Physics is an experimental science and requires observers.
Nevertheless it is important that such a procedure is GI in the following sense:
all the other physical observers must agree on the value of the
constructed observable, which means that on performing the calculation in
a different gauge (reference system) one obtains always the same result.
The procedure used to construct the observable is given
in a GI way.
In \cite{All, FMVV_II} several statements about quantum backreaction on the 
space-time expansion rate were given analyzing the expansion scalar, i.e.
a non GI quantity since it is a scalar field.
But this problem can be by-passed, see \cite{GMV1, GMV2}, introducing
averaged observables keeping into account consistently the volume measure.
Indeed  the vacuum expectation value of a scalar quantity,
as the expansion scalar, is still a gauge dependent value (on applying a
gauge transformation on such value one gets a different result).
On the other hand taking into account consistently the volume measure
one finally obtains a genuine GI averaged quantity.

Becoming a little more specific we first note that in \cite{GMV1,GMV2}
the class of non local observables considered is constructed averaging a 
scalar over a set of space-like hypersurfaces $\Sigma_A$ which belong to a foliation of the space-time over which another scalar field $A(x)$ takes constant value.
Therefore such a scalar field $A(x)$ characterizes a class of observers and the resulting averaged quantity is GI but observers dependent.
Consequently an analysis regarding the dynamics of the Universe must be associated to different observers and observables 
whose nature will be linked to the physical phenomena under consideration, and which can be then eventually probed by the measurements.
This is also a general theoretical problem and we find that a chaotic inflationary model with a non self-interacting massive inflaton is a very convenient playground to increase our understanding since the backreaction can drastically change for different observers.
As mentioned we started to investigate in this direction in~\cite{FMVVprl}.

In this framework the perturbative approach (for a useful formalism in gravity see~\cite{MetAll}) is the most viable way to perform computations. It means that one expands formally any quantity $Q=Q^{(0)}+Q^{(1)}+Q^{(2)}+\cdots$. The equations are then splitted in a set of equations which permit to find the dynamical quantities Q at various order in perturbation theory. This can be done in any system of reference and it is well known how such quantities transform changing the reference system. In our case we started by the solutions obtained long ago in the UCG \cite{FMVV_II}. Since our study is perturbative up to second order in the  fluctuations, one has to allow, in a general reference system, the field A defining the class of observers to be decomposed in components of different perturbative order $A=A^{(0)}+A^{(1)}+A^{(2)}+\cdots$.

In general to choose a gauge means to select the parameterization of the inhomogeneous degrees of freedom to be used to characterize the problem under investigation.
Among the possible choices of gauge, for a  single field driven inflationary phase, the most popular are the ones which kill two scalar fluctuations (to the desired order in perturbation theory) 
among the ones of the inflaton field and of the metric in the standard decomposition. 
We shall restrict ourself to the class of observers corresponding to such gauges, 
namely the scalar fields which characterize our observers will be the ones homogeneous in one of these gauges
(see \cite{Mar} for details about this correspondence).
We have already shown in \cite{FMVVprl} that, in the 
leading order in the slow-roll parameter and in the long wavelength limit, 
the observers which always  see as homogeneous the inflaton field are equivalent to the geodetic 
(free-falling) ones, up to second order in the cosmological perturbations. 
We have also shown that in the same level of approximations for such observers one cannot find any trace of a quantum backreaction on the effective Hubble factor and equation of state. 
Moreover we have also underlined \cite{FMVVprl} as the observers associated with a scalar homogeneous in the gauge with no perturbation on the spatial part of the metric see no backreaction.
One can go one step further and see that all the other choices, except one, are indeed associated to measurement performed by one  of the classes of observers introduced above, up to second order in perturbation theory and in the approximations considered, and correspond to the case wherein there is no leading backreaction in the observables we consider.

We are therefore left with only one last simple case which is the one 
defined adopting a foliation homogeneous
in the longitudinal gauge, where all the non zero scalar metric fluctuations are in the diagonal part of the metric. The corresponding class of observers, which sees an inhomogeneous and isotropic Universe, was shown to be non geodetic~\cite{FMVVprl} already at first order in the cosmological perturbations.
In this manuscript we analyze in detail this case, associated to a non free-falling class of observers which sees
an isotropic Universe and we will call such observers, as a consequence, isotropic observers.
In particular, we focus on the GI backreaction of the scalar cosmological fluctuations, in the long wavelength limit and at the lowest order in the slow-roll parameter, on the effective Hubble factor and equation of state.

In Section $2$ we review the main framework on which our work is based.
Then in Section $3$ we describe and characterize our observers (and, as a consequence, our observables) and in Section $4$ we study the backreaction problem associated to these isotropic observers computing the GI quantum corrections to the above mentioned observables.
Finally in the subsequent Section we give our conclusion. Two appendix follows, the first one with 
the general gauge transformation which connects different gauges, while the second one with
some details related to the computation of the average of non local operators which appear in this context.  

\section{Gauge invariant backreaction}
The advantages of a fully gauge invariant framework are well known.
We follow here a recent proposal~\cite{GMV1,GMV2} in this direction and recall the main tools we are going to use in our analysis. 
The main reason is that this formulation is suitable to let us define observables, of a non local nature
and constructed with quantum averages, which obey GI dynamical equations. Essentially the results of these investigations have been to give a prescription on how to construct and study a classical or quantum GI average of a scalar $S(x)$ (a classical field or a composite quantum operator).
This task has been achieved considering an hypersurface, which defines a class of observers,
with respect to which the average is done. 
A practical way to construct a suitable hypersurface, which we shall denote $\Sigma_{A_0}$,
consists in introducing another scalar field $A(x)$, with a time-like gradient, and then impose the constraint $A(x)=A_0$,
with $A_0$ a constant.
In this way even if both the scalars $S$ and $A$ are not GI, one can easily construct
an average which does have the desired property.
As it is reasonable in an inflationary context we shall consider here a spatially unbounded $\Sigma_{A_0}$. 
This is a consequence of the fact that expectation values of quantum 
operators can be extensively interpreted (and re-written) as spatial integrals 
weighted by the integration volume $V$, according to the general prescription \cite{GMV1}
\beq
\l \dots \re  ~~~\ra ~~~ V^{-1} \int_V d^3 x \left( \dots\right)\,,
\nonumber
\eeq
where the integration volume extends to all three-dimensional space.

Let us also observe that, in investigations such as the actual dynamics of the Universe,
one may imagine more suitable choices  to prove the phenomenological reconstruction of the space-time metric
based on past light-cone observations. A first step in this direction has been done in \cite{GMNV} where a covariant and GI 
formalism to average over the past light-cone of a generic observer was given.
 
As a consequence of its gauge invariance, an averaged quantity (observable) defined in the way
mentioned above can be computed in any gauge (coordinate frame).
Thanks to this freedom let us now consider a specific way of defining such a quantum 
(see \cite{GMV1,GMV2} for the classical counterpart) 
averaging prescription of a scalar quantity $S(x)$ as a functional of $A(x)$ thinking about how this can be reduced
in the (barred) coordinate system $\bar{x}^{\mu}=(\bar{t}, \vec{x})$ where the scalar
$A$ is homogeneous.
In such a reference frame one writes the simple expression~\cite{GMV1,GMV2}  
\beq
\langle S \rangle_{\!A_0}={\langle \sqrt{|\overline{\gamma}(t_0, {\vec{x}})|} 
\,~ \overline{S}(t_0, {\vec{x}}) \re \over  \langle   
\sqrt{|\overline{\gamma}(t_0, {\vec{x}})|} \re},
\label{media}
\eeq
where we have called $t_0$ the time $\bar{t}$ when $\bar{A}(\bar{x})=A^{(0)}(\bar{t})=A_0$.
In our notation  $\sqrt{|\overline{\gamma}(t_0, {\vec{x}})|}$ is the square root of the
determinant of the induced three dimensional metric on $\Sigma_{A_0}$.
It is also convenient to define the unit vector $n^\mu$ normal to the hypersurface $\Sigma_A$, which characterizes the physical  properties of the observers, as
\be
n^\mu= -Z_A^{-1/2} \partial^\mu A \quad , \quad
Z_A=-\partial^\mu A\partial_\mu A\,.
\label{gradA}
\ee

In this framework the dynamics of a perfect fluid-dominated early Universe can be conveniently described
by an effective scale factor $a_{\eff}=\langle\sqrt{|\bar{\gamma}|}\, \rangle ^{1/3}$ (where we have chosen $A^{(0)}(t)=t$ to 
have standard results at the homogeneous level \cite{Mar})
and we can write a quantum gauge invariant version of the effective
cosmological equations for the averaged geometry as \cite{GMV2}

\bea
&{}&\!\!\!\!\!\!\!\left(\frac{1}{a_{\eff}}\frac{\partial \, a_{\eff}}{\partial A_0} \right)^{\!2}=\nonumber \\
&{}& \hspace{-1cm}
\frac{8\pi G}{3} \left\langle\frac{\varepsilon}{Z_A}
\right\rangle_{\!\!\!\!A_0}\!\!
-\frac{1}{6}
\left\langle\frac{{\cal R}_s}{Z_A} \right\rangle_{\!\!\!\!A_0}\!\!
-\frac{1}{9}\left[   
\left\langle\frac{\Theta^2}{Z_A} \right\rangle_{\!\!\!\!A_0}\!\!\!
-\left\langle\frac{\Theta}{Z_A^{1/2}} 
\right\rangle_{\!\!\!\!A_0}^{\!\!2} \right]\!\!
+\frac{1}{3} \left\langle\frac{\sigma^2}{Z_A} \right\rangle_{\!\!\!\!A_0}\!
=\frac{1}{9} \left\langle\frac{\Theta}{Z_A^{1/2}} 
\right\rangle_{\!\!\!\!A_0}^{\!\!2},
\label{EQB1}\\
&{}&\!\!\!\!-\frac{1}{a_{\eff}} \frac{\partial^2 \, a_{\eff}}{\partial A_0^2}=\nonumber \\
&{}&\hspace{-1cm}\frac{4 \pi G}{3} \left\langle\!\frac{\varepsilon+3 {\pi}}{Z_A} 
\!\right\rangle_{\!\!\!\!A_0}\!\!\!\!-\frac{1}{3}\left\langle
\frac{\nabla^\nu(n^\mu\nabla_\mu n_\nu)}{Z_A} \right\rangle_{\!\!\!\!A_0}
\!\!\!\!+\frac{1}{6} \left\langle\!\frac{n_\mu
\partial^\mu Z_A}{Z_A^{2}} \Theta \!\right\rangle_{\!\!\!\!A_0}
\!\!\!\!-\frac{2}{9}\!\left[   
\left\langle\!\frac{\Theta^2}{Z_A} \!\right\rangle_{\!\!\!\!A_0}\!\!\!
\!-\left\langle\!\frac{\Theta}{Z_A^{1/2}} 
\!\right\rangle_{\!\!\!\!A_0}^{\!\!2} \right]\!\!
+\frac{2}{3} \left\langle\!\frac{\sigma^2}{Z_A}\! \right\rangle_{\!\!\!\!A_0}\,.
\label{EQB2}
\eea
In these equations several quantities appear (see \cite{GMV2} for details).
${\cal R}_s$ is a generalization of the intrinsic scalar 
curvature, $\Theta=\nabla_\mu n^\mu$
the expansion scalar, $\sigma^2$ the shear scalar, with respect to the
observers, and
\begin{eqnarray}
\varepsilon &=& \rho - (\rho+p)\left(1- (u^\mu n_\mu)^2 \right)\; , \
\\ 
\pi &=& p - \frac13 (\rho+p) \left(1- (u^\mu n_\mu)^2\right)\;,
\end{eqnarray}
with $u_\mu$ the 4-velocity comoving with the perfect fluid
and $\rho$ and $p$, respectively, the (scalar) energy density and 
pressure in the fluid's rest frame. In the inflationary scenario 
the fluid is given by the inflaton field.
Moreover we define the effective observers dependent energy density $\rho_{\eff}$
by writing the r.h.s. of Eq. \eqref{EQB1}
as $(8\pi G/3) \rho_{\eff}$ while the effective pressure $p_{\eff}$ 
is obtained by rewriting the r.h.s. of Eq. \eqref{EQB2} as
$(4\pi G/3) (\rho_{\eff}+3\, p_{eff})$. 

In the above set of equations in order to arrive to the covariant generalization of the second effective cosmological equation (\ref{EQB2}) we can start from the simple relation
\beq
{1\over a_{\eff}}{\pa^2 a_{\eff} \over \pa A_0^2}= 
{\pa \over \pa A_0}\left({1\over a_{\eff}}{\pa a_{\eff} \over \pa A_0}\right)+
\left({1\over a_{\eff}}{\pa a_{\eff} \over \pa A_0}\right)^2 
\label{317}
\eeq
and apply this to the following relation (see Eq.(\ref{EQB1}))
\be 
\frac{1}{a_{\eff}} \frac{\partial \, a_{\eff}}{\partial A_0}=\frac{1}{3} \left\langle\frac{\Theta}{(-\partial_{\mu} A \partial^{\mu}
A)^{1/2}} 
\right\rangle_{A_0}\,.
\label{HD}
\ee
Then, using the generalization of the Buchert-Ehlers commutation rule \cite{BE}~\footnote{Specifically using the generalized relation given in Eq. (3.10) of~\cite{GMV2}.}  and the Raychaudhuri's equation (see \cite{GMV2} for details), we obtain Eq.(\ref{EQB2}).
Taking an alternative point of view, starting from Eqs.(\ref{EQB1},\ref{EQB2}) and "imposing" their consistency, we are lead to a necessary condition of integrability (see \cite{Buchert}, or \cite{Larena} for a 
generalisation to non-matter comoving setups). 
Such integrability condition is often used to make general considerations and in our approach it is trivially satisfied.
Indeed, in the case under investigation (quantum backreaction of cosmological fluctuations in  an inflationary model) the dynamical equations for gravity and matter are solved order by order in perturbation theory and the solutions for the quantum fluctuations of both the scalar components of the metric and the inflaton field are explicitly derived in the long wavelength limit.
The different scalar constraints are, as a consequence, trivially satisfied order by order.
To conclude this remark we stress that the final value associated to the left hand side of Eq.(\ref{EQB2}) can be consistently evaluated, as it will be done in Section $4$, taking the derivative with respect to $A_0$ on the expression we obtain for the right hand side of Eq.(\ref{EQB1}).

We come back now to the definition of the shear with respect to the particular
class of observers defined by the scalar $A(x)$. 
To this purpose, we first introduce
the projector $h_{\mu\nu}$ into the tangent space of the hypersurfaces by \cite{GMV2}:
\begin{equation}
h_{\mu\nu} = g_{\mu\nu} + n_{\mu}n_{\nu}\,,~~~~~~~~~
h_{\mu\rho}h^{\rho}_{\nu} = h_{\mu\nu}\, , ~~~~~~~~~
h_{\mu\nu}n^{\mu} = 0\,.
\end{equation} 
Then considering the space-time flow generated by the time-like vector field $n_\mu$, we can 
define the expansion tensor of the flow worldlines as \cite{EvE}
\be
\Theta_{\mu\nu}\equiv 
h^\alpha_\mu h^\beta_\nu \nabla_\a n_\b
=\frac{1}{3} h_{\mu\nu}\Theta+\sigma_{\mu\nu}+\omega_{\mu\nu}\; ,
\ee
where
\be 
\sigma_{\mu\nu}\equiv h^\alpha_\mu
h^\beta_\nu \left(\nabla_{(\a} n_{\b)}-\frac{1}{3}h_{\alpha\beta}
\nabla_\tau n^\tau \right), ~~~ \omega_{\mu\nu}\equiv
h^\alpha_\mu h^\beta_\nu \nabla_{[\a} n_{\b]}
\ee
are the shear tensor and the rotation tensor, respectively.
In the case we are addressing, since $n_{\mu}$ is given by Eq. (\ref{gradA}), one has
a zero rotation tensor and can write 
\be
\sigma_{\mu\nu}=\Theta_{\mu\nu}-\frac{1}{3} h_{\mu\nu}\Theta \, , ~~~~~~~ \sigma^2\equiv\frac{1}{2} \sigma^\mu_\nu \sigma^\nu_\mu=
\frac{1}{2} \left(\Theta^\mu_\nu \Theta^\nu_\mu-\frac{1}{3}
\Theta^2\right)\, .
\label{sigma2}
\ee

Having to deal with the metric components in any specific frame, we employ
the standard decomposition of the metric in terms of scalar, transverse vector 
($B_i$, $\chi_i$) and traceless transverse tensor ($h_{ij}$) 
fluctuations up to the second order
around a homogeneous FLRW space-time
\bea
g_{00}\!\!&=&\!\! -1\!-\!2 \a\!-\!2 \a^{(2)}\,, \,\,  
g_{i0}=-{a\over2}\!\left(\beta_{,i}\!+\!B_i \right) \!
-\!{a\over2}\!\left(\!\beta^{(2)}_{,i}\!+\!B^{(2)}_i\!\right) 
\nonumber\\
\!\!\!\!g_{ij}\!\! &=&\!\!  a^2 \!\Bigl[ \delta_{ij} \! 
\left( \!1\!-\!2 \psi\!-\!2 \psi^{(2)}\right)
+D_{ij} (E+E^{(2)})
\nonumber\\
& & \!\!\!\!\!\!\!\!\!\!
+{1\over 2} \left(\chi_{i,j}+\chi_{j,i}+h_{ij}\right)
+ {1\over 2} \left(\chi^{(2)}_{i,j}+\chi^{(2)}_{j,i}+h^{(2)}_{ij}\right)\Bigr]\,,
\label{GeneralGauge}
\eea
where we have introduced the operator $D_{ij}=\partial_i\partial_j-\delta_{ij}\nabla^2/3$.
In the above expressions for notational simplicity we have removed an upper script for first order quantities.
Obviously the Einstein equations connect those fluctuations with the matter ones and
in particular we need to write also the inflaton field in terms of perturbations up to second order as
$\Phi(x)=\phi(t)+\varphi(x)+\varphi^{(2)}(x)$.

For such a perturbed FLRW space-time the shear scalar, with respect to the observers
defined by the scalar $A(x)$, is given, up to second order (hereafter we shall neglect vector contribution, which die away kinematically), by
\begin{eqnarray}
& & (\sigma^2)^{(0)}=0 \,\,\,\,\,,\,\,\,\,\, (\sigma^2)^{(1)}=0 \label{ShearUp1}\\
& & (\sigma^2)^{(2)}=\frac{1}{2 a^4 \dot{A}^{(0)\,2}}\left[A^{(1)}_{,i j}
A^{(1),i j}-\frac{1}{3}(\nabla^2 A^{(1)})^2 \right]
+\frac{1}{8 a^2}\left[\beta_{,i j}
\beta^{,i j}-\frac{1}{3}(\nabla^2 \beta)^2 \right]\nonumber\\
& & -\frac{1}{2 a^3 \dot{A}^{(0)}}\left[A^{(1)}_{,i j}
\beta^{,i j}-\frac{1}{3}(\nabla^2 A^{(1)}) (\nabla^2 \beta) \right]
-\frac{1}{4 a^2 \dot{A}^{(0)}} A^{(1)}_{,i j} \dot{\hat{h}}^{i
j}+\frac{1}{8 a}  \beta_{,i j} \dot{\hat{h}}^{i j}+\frac{1}{32}
\dot{\hat{h}}_{i j} \dot{\hat{h}}^{i j}
\label{ShearUp2}
\end{eqnarray}
where $\hat{h}_{i j}=2 D_{i j} E + h_{i j}$

These general perturbed expressions can be gauge fixed. Let us recall
some of the most common gauge fixing  of the scalar part of the metric
(which, in the same way, can be extended to all order in 
perturbation theory and are taken up to second order in our case):
the synchronous gauge (SG) is defined by $\alpha=0$ and $\beta=0$,
the uniform field gauge apart from setting $\Phi(x)=\phi(t)$ must be
supplemented by other conditions (for example $\beta=0$),
the uniform curvature gauge (UCG) is defined by
$\psi=0$ and $E=0$, finally the longitudinal gauge by $\beta=0$ and $E=0$. 

In order to deal with the observables and study Eqs.~(\ref{EQB1},\ref{EQB2})
one should study the dynamics of the inflaton and metric fluctuations, up to second order.
As a consequence of the gauge invariance of the observables considered, 
such solutions  can be computed in any frame convenient for the calculations.

In order to relate scalar fluctuations defined in different gauges, one has to give
the corresponding coordinate transformations up to second order (see Appendix A).
Using such coordinate transformations we can define the expression associated with Eq.(\ref{media}),
namely go from a general coordinate system to the barred one.

Let us consider the most important observable in our investigation, the average expansion scalar
which appears in Eq. \eqref{EQB1}.   We restrict ourselves to the long wavelength limit approximation. 
In this limit all the partial quantities we need can be written as follows: the expansion scalar can be written in the form
\be
\bar{\Theta}=3H-3 H \bar{\alpha}-3\dot{\bar{\psi}}
+\frac{9}{2} H \bar{\alpha}^2+3 \bar{\alpha} \dot{\bar{\psi}}
-6\bar{\psi}\dot{\bar {\psi}} -3 H \bar{\alpha}^{(2)}-
3 \dot{\bar{\psi}}^{(2)}-\frac{1}{8}h_{ij}\dot{h}^{ij}
\label{Theta} \,,
\ee
moreover the norm to construct the unit vector is
\be
- \partial_\mu \bar{A} \partial^\mu \bar{A}= 1- 2 \bar{\alpha}+4 
\bar{\alpha}^2- 2 \bar{\alpha}^{(2)}\,,
\label{partA}
\ee
while the measure in the spatial section reads
\be 
\sqrt{|\bar{\gamma}|}=a^3 \left(1-3 \bar{\psi}+\frac{3}{2}\bar{\psi}^2
-\frac{1}{16} h^{ij}h_{ij}
-3\bar{\psi}^{(2)}\right)\,.
\label{detgamma}
\ee 
Hereafter we shall neglect in our computations the dependence on the tensor fluctuations (which in the model that we are going to consider are negligible with respect to the scalar ones).
Collecting the results of Eqs.(\ref{Theta}, \ref{partA}, \ref{detgamma}) and inserting them in
Eq.(\ref{EQB1})
one obtains up to second order the simple expression
\be
H_{\eff}^2\equiv \left(\frac{1}{a_{\eff}}\frac{\partial \, a_{\eff}}{\partial A_0} \right)^2 = 
H^2 \left[1+\frac{2}{H}\langle \bar{\psi}\dot{\bar{\psi}}\rangle-
\frac{2}{H}\langle \dot{\bar {\psi}}^{(2)}\rangle\right]\,.
\label{Heff}
\ee
This general formula is the starting point of our analsys.

\section{Isotropic observers}
As we have already discussed we found useful in \cite{FMVVprl} to characterize the observers as free falling or not,
up to the second order in perturbation theory.
The free falling kinematics for an observer with velocity $v_\mu$ is determined by the equation $t_\mu=v^\nu \nabla_\nu v_\mu=0$ 
which can be determined in any reference frame from the corresponding metric.
Since in our analysis we keep contributions up to second order in the fluctuations we consider the following decomposition
$v_\mu=v_\mu^{(0)}+ v_\mu^{(1)} +v_\mu^{(2)}$.

Our first task is to characterize the scalar field $A(x)$ with the observers velocity $n_\mu$
which is associated with this class of free-falling observers within the limit of the approximations considered.
Let us exhibit the general condition at first order, since the zero order condition is trivially satisfied for any scalar.
In this case $t_\mu=n^\nu \nabla_\nu n_\mu$ should be zero at the first order:
for $\mu=0$ this condition is always satisfied while for $\mu=i$ one has
\be 
\frac{d}{dt}\left(\frac{A^{(1)}}{\dot{A}^{(0)}}\right)-\alpha=0\,.
\label{Cond_Geodesic}
\ee
Let us immediately emphasize that this property is not valid in general for all observers in the long wavelength limit.

We proceed now to define the class of observers which plays the central role in our investigation. 
We will call it isotropic observers and it is the one that sees an inhomogeneous and isotropic space.
From an intuitive point of view it is easy to see that it is defined by the property of the scalar field $A(x)$ being homogeneous in the longitudinal gauge. 
Such a scalar $A(x)$ is defined to first order by \cite{FMVVprl,Mar}
\be
A(x)=A^{(0)}+\dot{A}^{(0)} \left[\frac{a}{2}\beta+\frac{a^2}{2}\dot{E}\right]\,.
\label{A1_LG}
\ee
This is sufficient to see, from Eq.(\ref{Cond_Geodesic}), that, as recognized in \cite{FMVVprl}, these observers are not free-falling.

As said,
the longitudinal gauge is associated with a coordinate frame with non zero scalar fluctuations only on the diagonal part of the metric
and, as we can see from Eqs.(\ref{ShearUp1},\ref{ShearUp2}), it gives rise to a null shear scalar up to second order in perturbation theory
($A^{(1)}=\beta=E=0$) since we also have chosen to neglect the tensor contributions which are subleading with respect to the scalar ones. 
It is then easy to see that a direct consequences of $\sigma^2=0$ is that the shear tensor $\sigma_{\mu\nu}$ is also zero (again up to second order in perturbation theory).
So for this particular case we have
\be
\Theta_{\mu\nu}
=\frac{1}{3} h_{\mu\nu}\Theta\; ,
\ee
and the expansion is seen as isotropic from all the associated observers (see, for example, \cite{PeterUzan}).

As shown in \cite{Mar} the physical properties of a class of  observers associated with a given scalar are independent from the gauge considered. In particular this is the case also for the isotropic observers: in fact substituting Eq.~(\ref{A1_LG}) in Eq.~(\ref{ShearUp2}) one can see that the shear scalar $\sigma^2$ becomes identically zero independently from the gauge chosen.

We stress that, on defining our observers through a scalar field  $A$ homogeneous in a particular gauge (where we limit ourself to the case where the gauge is defined in the same way to all order in perturbation theory), we have a correspondence between a class of gauge choices and a class of observers with their physical properties (see \cite{Mar}). 

In the long wavelength limit such physical properties are characterized by the time gauge condition 
on $\epsilon^0_{(1)}$ and $\epsilon^0_{(2)}$ (see appendix A).
The ones which correspond  to the gauges with $\alpha=0$ are free-falling, the ones 
with $\psi=0$ have trivially zero backreaction (see Eq.(\ref{Heff})), while also the ones with $\varphi=0$ can be shown to be free-falling (see \cite{FMVVprl}). We are therefore left  with only one other case, with the time gauge condition given by  $\beta=0$ and $E=0$, which is precisely characterizing the isotropic observers.


\section{Backreaction for the isotropic observers}

Let us consider a spatially flat Universe, whose dynamics is driven 
by a minimally coupled scalar field, described by the action 
    \be
    S = \int d^4x \sqrt{-g} \left[ 
\frac{R}{16{\pi}G}
    - \frac{1}{2} g^{\mu \nu}
    \partial_{\mu} \phi \partial_{\nu} \phi
    - V(\phi) \right]\ .
    \label{action}
\ee
with potential $V(\phi)=1/2 \,m^2 \phi^2$.
The Friedmann equation for such a potential is given by:
\be
H^2 = 
\frac{1}{3 M_{\rm pl}^{2}} \left[ \frac{\dot\phi^2}{2} + m^2
\frac{\phi^2}{2} \right] \simeq \frac{m^2}{M_{\rm pl}^{2}}
\frac{\phi^2}{6}\,,
\ee
where $8 \pi G = M_{\rm pl}^{-2}$ and we consider slow-roll approximation.
Again in the slow-roll approximation, the inflationary trajectory 
is well approximated by:
\begin{eqnarray} \label{ansatz2}
H&=&H (t) \simeq H_0 - \frac{m^2}{3} (t-t_0)
\\
\phi (t) &\simeq& \phi_0 - \sqrt{\frac{2}{3}} m M_{\rm pl} (t - t_0) \\
a(t) 
&\simeq& a_0 \exp \left[ \frac{3}{2 m^2} \left( H_0^2 - H^2 \right) \right]\,,
\end{eqnarray}
where, hereafter, the subscript $0$ denotes the beginning of inflation.
In this single field inflationary model,
the scalar sector, scalar perturbations of the metric (Eq.(\ref{GeneralGauge})) and of the inflaton field,
have only one independent degree of freedom, which can be studied by
means of a single gauge invariant variable, such as the so-called
Mukhanov variable $Q$ \cite{Mukhanov}.
Before fixing the gauge such GI variable is given to first order by
\be 
Q=\varphi+\frac{\dot{\phi}}{H}\left(\psi+\frac{1}{6} \nabla^2 E \right)
\label{Mukhanov_variable}
\ee
and it always satisfies the same equation of motion independently from the gauge
\be
\ddot{Q} + 3 H \dot{Q} - \frac{1}{a^2}\nabla^2 Q +
\left[ V_{\phi \phi} + 2 \frac{d}{dt}\left(3 H + 
\frac{\dot H}{H}\right)\right] Q = 0 \ ,
\label{Eq_mukhanov}
\ee
with long wavelength solution $Q=f(\vec{x}) \frac{\dot{\phi}}{H}$.
Following \cite{FMVV_II} we can calculate, in the long wavelength limit and at the leading order in the slow-roll
parameter $\epsilon=-\dot{H}/H^2$, the renormalized value for the correspondent correlator. 
In particular, for $H_0\gg H$, one finds the well known result
\be
 \frac{\langle Q^2\rangle}{M_{pl}^2}\simeq-\frac{1}{24\pi^2} 
\frac{H^6_0}{M_{pl}^2 H^2 \dot{H}}  \sim \frac{H_0^4}{H^2 M_{pl}^2} \ln a
\label{phiquad}
\ee
(see~\cite{starall2} for a generic single field inflationary scenario).

Let us also put in evidence as in the UCG (which is the gauge considered in \cite{FMVV_II}) the
Mukhanov variable coincides with the inflaton perturbation to all orders. 
For such a gauge we have the following useful relation to first order
\begin{eqnarray}
\frac{H}{a} \nabla^2 \beta &=& 8 \pi G \left ( \dot{\phi}
\dot{\varphi} + V_\phi \varphi + 2 \, V \alpha \right)  
= 8 \pi G \frac{\dot{\phi}^2}{H} \, \frac{d}{dt} \left
(\frac{H}{\dot{\phi}}\varphi \right )\,,
\label{Eq_beta} \\
\alpha &=& 4 \pi G \frac{\dot{\phi}}{H} \, \varphi
\label{Eq_alpha} 
\end{eqnarray}
and we also have the relation $\dot \beta + 2 
H \beta = 2 \alpha / a$ (replacing the equality 
$\alpha = \psi$ in the longitudinal gauge). 

We shall also need the following relation for the second order perturbation of $\alpha^{(2)}$, 
valid in the long wavelength limit and at the leading order in the slow-roll parameter,

\begin{eqnarray}
\langle \alpha^{(2)} \rangle = \frac{1}{M_{pl}^2}\,\epsilon \,
 \langle \varphi^2 \rangle \,.
 \label{avealpha2}
\end{eqnarray}

We want now to determine the backreaction effects experienced by 
a class of isotropic observers in this chaotic inflationary model.
As a first step we need to compute the dynamical evolution of the scalar perturbations, at first and second order, in the longitudinal gauge (the barred gauge for this case).
As said, in doing this we shall always neglect vector and genuine tensorial contributions. 
Considering the gauge invariance of our procedure, 
we find convenient to use our previous results obtained in the UCG~\cite{FMVV_II}, which we have briefly recalled above, 
and express the barred quantities as functions of the scalar perturbations in the UCG.
In order to do that we shall find the proper gauge transformations to go from UCG to the longitudinal gauge. The general trasformations up to second order induced on a scalar field and on the metric are well known (e.g. see \cite{Mar}) and for
completeness are given in Appendix A. 
We first consider the quantities at first order and define some convenient variables which will be used throughout all our 
calculations also for the second order expressions.

\subsection{Longitudinal gauge quantities at first order}
Let us define $\hp=\frac{H \varphi}{\dot{\phi}}$ which, in UCG, is constant (in time) in the long wavelength limit.
In such a gauge we also have the following relations:
\be
\frac{a H}{2} \beta=r \, \hp \quad , \quad \alpha=-\frac{\dot H}{H^2} \hat{\psi}
\label{UCG1}
\ee

with $r$ is given by
\be
r=1-\frac{H}{a}\!\int\!\! dt\, a =-\frac{\dot H}{H^2}+{\cal O}\left(\frac{\dot{H}^2}{H^4}\right)\,,
\ee
where the first equality is obtained integrating the equation $\dot{\beta}+2H \beta=2\alpha/a$ in the long wavelength limit, while the second one comes from Eq.(\ref{Eq_alpha}).

Restricting ourselves to the scalar sector the transformation to go from the UCG to the longitudinal gauge 
is defined to first order by the two parameters $\epsilon^{0}_{(1)}$ and  $\epsilon_{(1)}$ (see Appendix A).
On requiring $\bar{\beta}=\bar{E}=0$ one obtains
\be
\epsilon^{0}_{(1)}=\frac{a}{2}\beta+\frac{a^2}{2}\dot{E}=\frac{a}{2}\beta \quad , \quad
\epsilon_{(1)}=\frac{E}{2}=0
\label{eps10}
\ee
and therefore we find
\be
\bar{\alpha}=\alpha-\frac{d}{dt}\left(\frac{a}{2}\beta\right) \quad , \quad
\bar{\psi}= \frac{a H}{2} \beta
\ee 
which are the non zero scalars which characterize the metric in the longitudinal gauge.

On using the relation \eqref{UCG1} one can easily check that
\be
\bar{\psi}=\bar{\alpha}=r\, \hat{\psi} \, ,
\ee
which satisfies the following constrain (obtained from the spatial off-diagonal Einstein equations)
\be
\dot{\bar{\psi}}+H \bar{\psi}=\frac{1}{2M_{pl}^2} \dot{\phi}\, \bar{\varphi} \,.
\ee
Here $\bar{\varphi}=\varphi \, (1-r)$ is the first order field perturbation in the longitudinal gauge, which is written in terms of the corresponding quantity in the UCG.
Note that
\be
\dot{r}=-H\left[r+\frac{\dot H}{H^2}(1-r)\right]
\ee
\subsection{Longitudinal gauge quantities at second order}
Let us compute again the scalar part of the coordinate transformation, this time at second order.
On requiring $\bar{E}^{(2)}=0$, after some algebra, we get (see Appendix A)
\be
\epsilon_{(2)}=\frac{3}{8} \frac{1}{\nabla^2} \left(\frac{\partial^i\partial^j}{\nabla^2}-\frac{1}{3}\delta^{ij}\right)
\partial_i \beta\, \partial_j \beta=\frac{3}{2}\frac{r^2}{a^2 H^2}\frac{1}{\nabla^2} \left(\frac{\partial^i\partial^j}{\nabla^2}-\frac{1}{3}\delta^{ij}\right)
\partial_i \hp\, \partial_j \hp\,,
\label{eps2s}
\ee

while on requiring $\bar{\beta}^{(2)}=0$ we obtain
\ba
\epsilon^0_{(2)}&=&a^2 \dot{\epsilon}_{(2)}+a \beta^{(2)}+\frac{1}{8} \frac{d}{dt}(a^2 \beta^2)+
\frac{a^2H}{4}\beta^2-a \alpha \beta - a \frac{\partial^i}{\nabla^2}(\alpha\, \partial_i \beta)\nonumber\\
&=& a^2 \dot{\epsilon}_{(2)}+a \beta^{(2)}-\frac{a}{2} \alpha \beta- a \frac{\partial^i}{\nabla^2}(\alpha\, \partial_i \beta)\,.
\label{eps20}
\ea
For our purposes it is sufficient to compute
the general expression for $\bar{\psi}^{(2)}$, which in the long wavelength limit approximation is given by 
\be
\bar{\psi}^{(2)}=\psi^{(2)}+\frac{H}{2} \epsilon^0_{(2)}-\epsilon^{0}_{(1)}\left(2H \psi^{(1)}+\dot{\psi}^{(1)}\right)-\frac{H}{2}
\epsilon^0_{(1)}\dot{\epsilon}^0_{(1)}-\left(H^2+\frac{\dot{H}}{2}\right)\left(\epsilon^{0}_{(1)}\right)^2\,.
 \ee
Substituting the results given in Eqs. \eqref{eps10} and \eqref{eps20}, and the fact that in the UCG $\psi=0$ and 
$\psi^{(2)}=0$, one finds the following expression
\be
\bar{\psi}^{(2)}=\frac{H a}{2}\left[a \dot{\epsilon}_{(2)}+\beta^{(2)}-\frac{\partial^i}{\nabla^2}(\alpha\, \partial_i \beta)\right]-
\frac{a H}{2} \alpha \beta-\frac{a^2}{8}\left(\dot H+H^2\right)\beta^2\,,
\ee
which, in terms of $\hp$, reads
\be
\bar{\psi}^{(2)}=\frac{H a}{2}\left[a \dot{\epsilon}_{(2)}+\beta^{(2)}\right]+
\left[\frac{3}{2} \frac{\dot H}{H^2} -\frac{1}{2}\left(1+\frac{\dot H}{H^2}\right)r\right]r \,\hp^2\,.
\label{psi2}
\ee

Such expression still depends on $\beta^{(2)}$ and this dependence can be exploited using the following 
second order relation
\be
\partial_i \partial_j \left[ \frac{1}{2a}\dot{\beta}^{(2)}+\frac{H}{a} \beta^{(2)}-\frac{1}{a^2}\alpha^{(2)}\right]+ F_{ij}=0\,,
\ee
with
\bea
F_{ij}&=&\frac{1}{a^2}\partial_i \alpha \,\partial_j \alpha -\frac{H}{a}\partial_i \beta \,\partial_j \alpha  -\frac{1}{2a} \dot \alpha \partial_i\partial_j \beta+
\frac{1}{4a^2}\left(\nabla^2 \beta \,\partial_i\partial_j \beta-\partial_i\partial_k \beta \partial_j \partial^k \beta\right)
\nonumber\\
&& 
-\frac{1}{M_{pl}^2}\left(\frac{1}{a^2}\partial_i \varphi \,\partial_j \varphi-\frac{\dot \phi}{2a} \partial_i \beta \,\partial_j \varphi\right)\,,
\eea
which can be easily extracted from the off-diagonal spatial Einstein equations, given in \cite{FMVV_II} 
in the UCG, using the relation $\dot{\beta}+2 H \beta-2/a \alpha=0$.
Acting on both sides with the operator $D^{ij}=\partial^i\partial^j- \delta^{ij}/3 \nabla^2$ (contracting both indices) one finds
\be
\dot{\beta}^{(2)}+2 H \beta^{(2)}=\frac{2}{a}\alpha^{(2)}- 3 a \frac{1}{\nabla^2}
\left(\frac{\partial^i\partial^j}{\nabla^2} F_{ij} - \frac{\delta^{i j}}{3} F_{ij} \right)\,,
\ee
which can be integrated to give
\be
\frac{a}{2} \beta^{(2)}=\frac{1}{a}\left[ \int\! dt \,a\, \alpha^{(2)}-\frac{3}{2}
\int dt a^3
\frac{1}{\nabla^2}
\left(\frac{\partial^i\partial^j}{\nabla^2} - \frac{\delta^{i j}}{3}\right) F_{ij} \right]\,.
\label{beta2}
\ee
In the slow-roll approximation the vacuum average of the first term on the right hand side 
can be evaluated with the help
of Eq.~\eqref{avealpha2}, 
while for $F_{ij}$ one has, in the long wavelength approximation, the following relation in terms of $\hat{\psi}$,
\bea
F_{ij}=\frac{1}{a^2} 
\left[ 2\frac{\dot H}{H^2}+\frac{\dot{H}^2}{H^4}+r\left(\frac{\ddot H}{H^3}-
\frac{\dot{H}^2}{H^4}\right)\right]\partial_i \hp \partial_j \hp\,.
\eea

\subsection{Gauge invariant backreaction}
\label{four3}
We can now proceed to evaluate, for the isotropic observers, the backreaction on the effective Hubble factor induced by scalar fluctuation, given by~\eqref{Heff}. 
The first correction can be shown to be given by
\be
\frac{2}{H}\langle \bar{\psi}\dot{\bar{\psi}}\rangle=\left[2\frac{\dot{H}^2}{H^4}+{\cal O}\left(\frac{\dot{H}^3}{H^6}\right)\right] \frac{\langle \varphi^2 \rangle}{M_{pl}^2}
\ee
which is of second order in the slow-roll parameter.
Let us remark again that $\langle \varphi^2 \rangle$ is the vacuum quantum average of the square of the first order inflaton fluctuation, computed in the UCG, and that in such a gauge the inflaton fluctuations coincides with the locally gauge invariant Mukhanov variable \cite{Mukhanov}.

The second correction in Eq.~\eqref{Heff} is determined by the time derivative of $\bar{\psi}^{(2)}$, given in Eq.~\eqref{psi2}.
After evaluating all the terms contributing to it we find that it is given by
\be
-\frac{2}{H}\langle \dot{\bar{\psi}}^{(2)}\rangle=\left[\frac{3}{5}\frac{\dot{H}}{H^2}+{\cal O}\left(\frac{\dot{H}^2}{H^4}\right)\right] \frac{\langle \varphi^2 \rangle}{M_{pl}^2}\,.
\ee
The leading contribution in the latter expression comes from the term proportional to $\beta^{(2)}$ which appears in Eq.~\eqref{psi2},
and in particular from the last non local term which contributes in Eq.~\eqref{beta2},while all the other contributions are subleading. We give some details of its computation in the Appendix B. 

Therefore, collecting these results, one has 
\be
H_{\eff}^2=\left(\frac{1}{a_{\eff}}\frac{\partial a_{\eff}}{\partial A_0}\right)^2=H^2 \left[1
+\frac{3}{5} \frac{\dot{H}}{H^2}\frac{\langle \varphi^2 \rangle}{M_{pl}^2}
+{\cal O} \left(\frac{\dot{H}^2}{H^4}\right)
\frac{\langle \varphi^2 \rangle}{M_{pl}^2} \right]\,.
\label{hubblelg}
\ee
and we find that the observers associated to the longitudinal gauge foliation, which are not 
free-falling, but see an inhomogeneous isotropic space, experiences a backreaction such that $H^2_{\eff}<H^2$. 

A valuable information is also given by the effective equation of state which is defined with respect to such observers.
Therefore we study the quantity $w_{\eff}=p_{\eff}/\rho_{\eff}$.

We start from a relation~\cite{FMVVprl}  valid for any class of observers and slow-roll inflationary
models, going this time up to second order in the slow-roll parameter $\epsilon$.
Namely, consider for the effective Hubble factor the following expression
\be
\left(\frac{1}{a_{\eff}}\frac{\partial \, a_{\eff}}{\partial A_0}
\right)^2\!= \frac{8\pi G}{3} \rho_{eff}=
H^2\! \left[1\!+ \left( 
c \frac{\dot{H}}{H^2}+ d \frac{\dot{H}^2}{H^4} +
{\cal O} \left(\frac{\dot{H}^3}{H^6}\right)\right)
\frac{\langle \varphi^2 \rangle}{M_{pl}^2}\right]\,,
\ee
where $c$ and $d$ are parameters which encodes the possible non zero backreaction at first and second order in the slow-roll approximation (in our case $c=\frac{3}{5}$ and $d$ is not fixed).
Then, from the consistency between the effective equations for the averaged geometry 
(see Section 2), one obtains
\be
-\frac{1}{a_{\eff}} \frac{\partial^2 \, a_{\eff}}{\partial A_0^2}=   \frac{4\pi G}{3} \left(\rho_{eff}+3 p_{eff} \right)=
-\dot{H}\!-\!H^2\!-\!H^2 \!\left[c \frac{\dot{H}}{H^2} +
\left(d-\frac{3}{2} c \right) \frac{\dot{H}^2}{H^4}+c \frac{\ddot{H}}{H^3}
+{\cal O} \left(\frac{\dot{H}^3}{H^6}\right)\right]
\frac{\langle \varphi^2 \rangle}{M_{pl}^2}\,.
\nonumber
\ee
From this relations it is easy to see that the effective equation of state to the first
non trivial order is given by
\be
w_{\eff}=\frac{p_{\eff}}{\rho_{\eff}}=-1-\frac{2}{3} \frac{\dot{H}}{H^2}
+\left[\frac{5}{3} c \frac{\dot{H}^2}{H^4}-\frac{2}{3} c \frac{\ddot{H}}{H^3}+{\cal O} \left(\frac{\dot{H}^3}{H^6} \right)\right]
\frac{\langle \varphi^2 \rangle}{M_{pl}^2}\,,
\ee
where the $d$ dependence disappears in the leading order correction.
Therefore for the isotropic observers and a $m^2 \phi^2$ chaotic model we obtain 
the following result ($\ddot{H}/H^3$ is, for this case, of third order in the slow-roll parameter 
$\epsilon$):
\be
w_{\eff}=\frac{p_{\eff}}{\rho_{eff}}=-1+\frac{2}{3} \epsilon+\left[
 \epsilon^2+{\cal O} \left(\epsilon^{3} \right)\right]
\frac{\langle \varphi^2 \rangle}{M_{pl}^2}\,.
\label{eqstatelg}
\ee
The correction to $w_{\eff}$ is positive and therefore goes in
direction of a less de Sitter like equation of state (namely the new
equation of state is farer from a de Sitter like equation of state,
with respect to the initial background value with no correction from
the backreaction).
In fact $H^2_{\eff}$ is also negatively corrected and one can also see
and easily check that
the pressure (being negative) receive instead a positive correction.
Summarizing the result, these isotropic non free-falling observers see
a slightly smaller expansion rate, less de Sitter like.

\section{Conclusions}
The gauge invariant averaging procedure, which has been recently developed in~\cite{GMV1,GMV2},
has been applied, in the context of cosmological perturbation theory, to inflationary early 
cosmology to analyze the so-called backreaction effects.
This approach is free from ambiguities by construction and make it possible to study the dynamics of the Universe
considering a set of non local gauge invariant observables. With a suitable large enough set of such observables (measurements), which are observer dependent, one might hope to obtain sufficient insights of the corresponding physics.  
In a previous work~\cite{FMVVprl}  we have considered a chaotic inflationary model with a non interacting massive inflaton and we have addressed a specific set of observers. 
Within the approximations employed, that is in the leading order in the slow-roll parameter and in the long wavelength limit, they resulted all to 
belong to two different classes of observers. One which trivially has no backreaction effects (observers associated 
to the UCG), another one associated to the free-falling observers.
Also the latter ones have the property of not experiencing any leading backreaction in the effective Hubble factor nor in the equation of state of the expanding phase. 
An analysis able to go beyond the limit of these approximations, such as considering short wavelength fluctuations
or going at higher order in the perturbation theory or even beyond a perturbative evaluation, is hard to be performed.

In this work for the same model we have addressed the case of other observables/observers and have mainly concentrated our efforts to compute the backreaction for the so called isotropic observers, which were already recognized in our previous work to be not free-falling. 
This class of observers sees an inhomogeneous and isotropic Universe and is characterized by zero shear scalar
and tensor.
We have found that for such a case the quantum fluctuations contribute, at the leading order in the slow-roll parameter, to the effective Hubble factor. This contribution comes from averaging a non local operator in the spatial foliation. 
We have shown that the average procedure of the non-local terms coming from Eq.(\ref{beta2}) gives a quantity free of unusual infrared divergencies (see Appendix B), since cancellations take place.
After computing the effective Hubble factor, see Eq.~\eqref{hubblelg}, we have found that the backreaction is negative, i.e. such observers experience a smaller expansion rate induced by gauge invariant cosmological fluctuations analyzed to second order in perturbation theory.
Another object of interest is the equation of state of the expanding phase. We have obtained an effective equation of state, 
see Eq.~\eqref{eqstatelg}, which is affected by the backreaction in such a way to appear less de Sitter like. 
Summarizing, we observe that the backreaction is not zero for the observers with zero shear scalar and tensor.
This is a manifest result of the fact that the "isotropic" observers are not free-falling, but they are accelerated in order to see a "diagonal" metric, without off-diagonal scalar quantum fluctuations.
Note that one might still consider the following (gedanken) experiment: one could construct the observables
corresponding to the free-falling and isotropic observers and reproduce experimentally the two frameworks to measure them (which
of course include quantum fluctuations). In such a case one could measure the difference of their corresponding values
(which would be zero for a classical homogeneous and isotropic model)
and define a new observable on which, of course, all others spectator observers do agree.
An observable which is perfectly a physical one.

We have seen that one can deal with different observables, non local but gauge invariant, and the associated measurement can probe for some of them backreaction effects and for others no backreaction at all.
Of course there is nothing wrong with that, since the observables are observers dependent.
We can conclude that different observers are able to probe different features of the Universe dynamics since the outcome of the measurement is a gauge invariant quantity.

\vskip 0.3cm
\noindent
{\bf Acknowledgements}

We wish to thank Fabio Finelli and Giovanni Venturi for stimulating discussions. 
GM wish to thank Maurizio Gasperini, Dominik 
J. Schwarz, Jean-Philippe Uzan and Gabriele Veneziano for comments and discussions.

\appendix

\section{General gauge trasformation}

Let us give for completeness the general ``infinitesimal'' 
coordinate transformation used to connect two different gauges.
Following, for example, what stated in \cite{Mar} such transformation can be 
parametrized by the  first-order, $\ep_{(1)}^\mu$, 
and second-order, $\ep_{(2)}^\mu$,  generators as \cite{MetAll}
\beq
x^\mu \rightarrow \tilde{x}^\mu= x^\mu + \epsilon^\mu_{(1)} +\frac{1}{2}
\left(\epsilon^{\nu}_{(1)}\pa_\nu \epsilon^{\mu}_{(1)} + 
\epsilon^{\mu}_{(2)}\right) + \dots
\label{311}
\eeq
where
\beq
\ep_{(1)}^\mu= \left( \ep_{(1)}^0, \hat{\ep}_{(1)}^{\,i} 
\right), ~~~~~~~~~
\ep_{(2)}^\mu= \left( \ep_{(2)}^0, \hat{\ep}_{(2)}^{\,i} \right) 
\label{312}
\eeq
and we can explicitly separate the scalar and the pure transverse
vector part of $\hat{\ep}_{(1)}^i$, $\hat{\ep}_{(2)}^i$ as 
\be 
\hat{\ep}_{(1)}^{\,i}=\pa^i \ep_{(1)}+ \ep_{(1)}^i \,\,\,\,\,\,\,\,\,\,,
\,\,\,\,\,\,\,\,\,
\hat{\ep}_{(2)}^{\,i}=\pa^i \ep_{(2)}+ \ep_{(2)}^i\,.
\ee 
Under the associated gauge transformation 
(or local field reparameterization) - where, by definition, old
and new fields are evaluated at the same space-time point x - 
 a general tensor changes, to first and second 
order, as
\be
T^{(1)} \rightarrow \tilde{T}^{(1)}=  T^{(1)}-L_{\ep_{(1)}} T^{(0)}
\label{GGT1}
\ee
\be
T^{(2)} \rightarrow \tilde{T}^{(2)}=  T^{(2)}-L_{\ep_{(1)}} T^{(1)}
+\frac{1}{2}\left( L^2_{\ep_{(1)}} T^{(0)}-L_{\ep_{(2)}} T^{(0)}\right)
\label{GGT2}
\ee
where $L_{\ep_{(1,2)}}$ is the Lie derivative respect the vector 
$\ep^\mu_{(1,2)}$.
In particular the scalar metric perturbations change,  to first order, as
\be
\alphat = \alpha - \dot \epsilon^0_{(1)},
\label{415}
\ee
\be
\betat = \beta - \frac{2}{a} \epsilon^0_{(1)} +2 a \dot \epsilon_{(1)},
\label{416}
\ee
\be
\tilde{\psi} =\psi+ H \epsilon^0_{(1)}+\frac{1}{3} 
\nabla^2\epsilon_{(1)},
\label{417}
\ee
\be
\tilde{E}=E -2\epsilon_{(1)} \,.
\label{418}
\ee
and to second order, as
\bea
\alphat^{(2)} &=& \alpha^{(2)}-\frac{1}{2}\dot \epsilon^0_{(2)}
- \epsilon^0_{(1)} \dot{\alpha}-2 \alpha
\dot \epsilon^0_{(1)}+\frac{1}{2}
\epsilon^0_{(1)}\ddot \epsilon^0_{(1)} +\left(\dot 
\epsilon^0_{(1)}\right)^2-\hat{\ep}_{(1)}^{\,i} \alpha_{,i}
\nonumber \\ & & 
+\frac{1}{2}
\left(\hat{\ep}_{(1)}^{\,i} \epsilon^0_{(1),i}\right)^.
-\frac{a^2}{2}
\dot{\hat{\ep}}_{(1)}^{\,i}\dot{\hat{\ep}}_{(1)i}-\frac{a}{2}\left(\beta_{,i}+
B_i\right) 
\dot{\hat{\ep}}_{(1)}^{\,i},
\label{422}
\\
\tilde{\beta}^{(2)}&=& \beta^{(2)} -\frac{1}{a} \epsilon_{(2)}^0
+ a \dot{\epsilon}_{(2)} +\frac{1}{2a}\frac{d}{d t} (\epsilon_{(1)}^0)^2
+\frac{1}{a}\hat{\ep}_{(1)}^{\,i}
\epsilon^0_{(1),i}- \hat{\ep}_{(1)}^{\,i}\beta_{,i}
+\frac{\partial^i}{\nabla^2} \Upsilon_i\,,
\label{423}\\
\tilde{\psi}^{(2)} &=& \psi^{(2)}+
\frac{H}{2}\epsilon^0_{(2)}+\frac{1}{6}\nabla^2 \epsilon_{(2)}
-\epsilon_{(1)}^0 \left(2 H \psi+\dot{\psi}
\right) -
\frac{H}{2} \epsilon^0_{(1)} \dot \epsilon^0_{(1)} 
\nonumber \\
& &
-\frac{\epsilon^{0 \, 2}_{(1)}}{2} \left( \dot H + 2 H^2 \right) 
-\frac{H}{2}\epsilon^{0}_{(1),i}\hat{\ep}^{\,i}_{(1)}
-\psi_{,i} \hat{\ep}^{\,i}_{(1)}
-\frac{1}{6} \Pi^i_i\,,
\label{424}
\\
\tilde{E}^{(2)} &=& E^{(2)}- \epsilon_{(2)}
-\frac{1}{2}\frac{1}{\nabla^2}\Pi^i_i
+\frac{3}{2}\frac{\partial^i\partial^j}{(\nabla^2)^2}\Pi_{ij}\,,
\label{425}
\end{eqnarray}
where
\begin{eqnarray}
\Upsilon_i &=& \frac{2}{a} \epsilon_{(1),i}^0 \dot{\epsilon}_{(1)}^0
- \epsilon_{(1)}^0  (\dot{\beta}_{,i}+\dot{B}_i)-
H \epsilon_{(1)}^0 (\beta_{,i}+B_i) -\dot{\epsilon}_{(1)}^0 
(\beta_{,i}+B_i)  
\nonumber \\ & &
- \frac{4}{a} \alpha \epsilon_{(1),i}^0 
-2a \dot{\hat{\ep}}_{(1)i} \left( 2 H \epsilon^0_{(1)}+\frac{1}{2}
\dot{\epsilon}^0_{(1)}+2 \psi\right)-a \ddot{\hat{\ep}}_{(1)i}\epsilon^0_{(1)}
\nonumber \\ & &
-a\left(\hat{\ep}^{\,j}_{(1)}\dot{\hat{\ep}}_{(1)i,j}+2 
\hat{\ep}^{\,j}_{(1),i}\dot{\hat{\ep}}_{(1)j}
+\dot{\hat{\ep}}^{\,j}_{(1)}
\hat{\ep}_{(1)i,j} \right)- \hat{\ep}^{\,j}_{(1)}B_{i,j}
-\hat{\ep}^{\,j}_{(1),i}B_{j}
\nonumber \\ & &
-2 a \dot{\hat{\ep}}^{\,j}_{(1)} 
\left[-D_{ij} E-\frac{1}{2}\left(\chi_{i,j}+\chi_{j,i}-
h_{ij}\right)\right]
\end{eqnarray}
\begin{eqnarray}
\Pi_{i j}&=&-2 \epsilon^{0}_{(1)} D_{i j} \left(H E
+\frac{\dot{E}}{2}\right) 
-\hat{\ep}^{\,k}_{(1)} D_{i j} E_{,k}
-\hat{\ep}^{\,k}_{(1),j} D_{i k} E
-\hat{\ep}^{\,k}_{(1),i} D_{j k} E
\nonumber \\
& & 
+2 H \epsilon^{0}_{(1)} \left(\hat{\ep}_{(1)i,j}+\hat{\ep}_{(1)j,i}\right)
+2 \psi \left(\hat{\ep}_{(1)i,j}+\hat{\ep}_{(1)j,i}\right) 
-\epsilon^{0}_{(1)}\left[H \left( 
\chi_{i,j}+\chi_{j,i}+h_{ij}\right)
\right. \nonumber \\
& & \left.
+\frac{1}{2} \left(\dot{\chi}_{i,j}+\dot{\chi}_{j,i}
+\dot{h}_{ij}\right)\right]
+\frac{1}{2}\epsilon^{0}_{(1)}\left(\dot{\hat{\ep}}_{(1)i,j}
+\dot{\hat{\ep}}_{(1)j,i}\right)
-\frac{1}{a^2} \epsilon^{0}_{(1),i}\epsilon^{0}_{(1),j}
\nonumber \\
& & 
+\hat{\ep}_{(1)k,i}\hat{\ep}^{\,k}_{(1),j}
+\frac{1}{2}\hat{\ep}^{\,k}_{(1)}
\left(\hat{\ep}_{(1)i,j k}+\hat{\ep}_{(1)j,i k}\right)
-\frac{1}{2}\hat{\ep}^{\,k}_{(1)} \left(
\chi_{i,j k}+\chi_{j,i k}+
h_{i j,k}\right)
\nonumber \\
& & 
+\frac{1}{2}\left(\dot{\hat{\ep}}_{(1) i}\epsilon^{0}_{(1),j}+
\dot{\hat{\ep}}_{(1) j}\epsilon^{0}_{(1),i}\right)
+\frac{1}{2 a}\left(\beta_{,i}+B_i\right) \epsilon^{0}_{(1),j}
+\frac{1}{2 a}\left(\beta_{,j}+B_j\right) \epsilon^{0}_{(1),i}
\nonumber \\
& & 
+\frac{1}{2}\left(\hat{\ep}_{(1)j, k}\hat{\ep}^{\,k}_{(1),i}+
\hat{\ep}_{(1)i, k}\hat{\ep}^{\,k}_{(1),j}\right)
-\frac{1}{2}\left(\chi_{i,k}+\chi_{k,i}+h_{i k}\right) \hat{\ep}^{\,k}_{(1),j}
\nonumber \\
& & 
-\frac{1}{2}\left(\chi_{j,k}+\chi_{k,j}+h_{j k}\right) \hat{\ep}^{\,k}_{(1),i}\,.
\end{eqnarray}

For the vector metric perturbations one obtains, to first order
\be
\tilde{B}_i=B_i+2 a \dot{\ep}_{(1)i}
\ee
\be 
\tilde{\chi}_i=\chi_i-2 \ep_{(1)i}
\ee
and to second order
\be
\tilde{B}^{(2)}_i=B^{(2)}_i+a \dot{\ep}_{(2)i}-\frac{\partial_i \partial^j}{\nabla^2}
\Upsilon_j +\Upsilon_i
\ee
\be 
\tilde{\chi}^{(2)}_i=\chi^{(2)}_i-\ep_{(2)i}
-2 \frac{\partial_i\partial^j\partial^k}{(\nabla^2)^2}\Pi_{j k}
+2 \frac{\partial^j}{\nabla^2}\Pi_{i j}
\ee
The metric tensor perturbation is gauge invariant to first order
($\tilde{h}_{i j}=h_{i j}$)  
while to second order one obtains 
\be
\tilde{h}_{i j}^{(2)}=h_{i j}^{(2)}+2 \Pi_{i j}
+\left(\frac{\partial_i \partial_j}{\nabla^2}-\delta_{i j}\right) \Pi^k_k
+\left(\frac{\partial_i \partial_j}{\nabla^2}+\delta_{i j}\right) 
\frac{\partial^k \partial^l}{\nabla^2}\Pi_{k l}
-\frac{2}{\nabla^2}\left(\partial_i \partial^k \Pi_{j k}
+\partial_j \partial^k \Pi_{i k}\right)\,.
\ee

To conclude the associated gauge transformation of a scalar 
field $A(x)=A^{(0)}(t)+A^{(1)}(x)+A^{(2)}(x)$
is, to first order, 
\be
A^{(1)}~~ \rightarrow ~~\ti{A}^{(1)}=A^{(1)}-\ep_{(1)}^0 \dot{A}^{(0)},
\label{315}
\ee
and, to second order, 
\bea
A^{(2)} ~~\rightarrow &&~~\ti{A}^{(2)}=  A^{(2)}-\ep_{(1)}^0 \dot{A}^{(1)}
-\left(\ep_{(1)}^i+\partial^i \ep_{(1)}\right) \partial_i A^{(1)}
\nonumber \\ 
&&+\frac{1}{2}\left[\ep_{(1)}^0 
\partial_t (\ep_{(1)}^0 \dot{A}^{(0)})+ \left(\ep_{(1)}^i+\partial^i \ep_{(1)}\right)\partial_i
\ep_{(1)}^0 \dot{A}^{(0)}-\ep_{(2)}^0 \dot{A}^{(0)}\right]\,.
\label{316}
\eea

\section{Non-local vacuum expectation value}
As we have observed in Section \ref{four3}, the leading contribution to the backreaction to the effective Hubble factor 
is coming from the term proportional to 
$\dot{\bar{\psi}}^{(2)}$ and originating from the last term in $\beta^{(2)}$ (see Eq.~\eqref{beta2}). 
By inspection of Eq.~\eqref{beta2} we see
that we need to compute the following average of a non local operator:
\be
\left\langle \hat{O}^{ij} \partial_i \vf\, \partial_j \vf \right\rangle \equiv
\left\langle \frac{1}{\nabla^2} \left(\frac{\partial^i\partial^j}{\nabla^2}-\frac{1}{3}\delta^{ij}\right)
\partial_i \vf\, \partial_j \vf \right\rangle\,.
\ee
Considering the second quantized fluctuations
\be 
    \hat{\varphi} (t, {\bf x}) = \frac{1}{(2 \pi)^{3/2}} \int d^3{\bf k}
    \left[ \varphi_{k} (t) \, e^{i \vec{k} \cdot \vec{x}} \, \hat{b}_{\bf
    k} +  \varphi^*_{k} (t)e^{- i \vec{k} \cdot \vec{x}}
    \hat{b}^\dagger_{{\bf k}} \right]
    \label{quantumFourier}
    \ee
where the $\hat{b}_{\bf k}$ are time-independent Heisenberg operators
with the usual commutation relations:
 \be
    [\hat{b}_{\bf k}, \hat{b}_{{\bf k}'}] =
    [\hat{b}^\dagger_{\bf k}, \hat{b}^\dagger_{{\bf k}'}] = 0 \, \quad
    [\hat{b}_{\bf k}, \hat{b}^{\dagger}_{{\bf k}'}] = \delta^{(3)}
    ({\bf k} - {\bf k}') \,,
\ee
the v.e.v. can be written as
\begin{eqnarray}
\left\langle \hat{O}^{ij} \partial_i \vf\, \partial_j \vf \right\rangle= 
\!\!\int \!\frac{d^3 {\bf p}}{(2 \pi)^3} d^3 {\bf k} \, \varphi_p \, \varphi^*_k \, e^{i (\vec{p}-\vec{k}) \cdot \vec{x}} 
\left[ -\frac{(p^i-k^i)(p^j-k^j)}{|\vec{p}-\vec{k}|^4}+\frac{1}{3}
\frac{\delta^{i j}}{|\vec{p}-\vec{k}|^2}\right] p_i k_j \delta^{(3)} (\vec{p}-\vec{k})\,.
\end{eqnarray}

Let us now make the following change of variable $\vec{p}-\vec{k}=\vec{q}$ to obtain
\begin{eqnarray}
\left\langle \hat{O}^{ij} \partial_i \vf\, \partial_j \vf \right\rangle= 
\int \frac{d^3 {\bf p}}{(2 \pi)^3} d^3 {\bf q} \, \varphi_p \, \varphi^*_{|\vec{p}-\vec{q}|} \, e^{i \vec{q} \cdot \vec{x}} 
\left[ \frac{1}{3}\frac{2 (\vec{p} \cdot \vec{q})+|\vec{p}|^2}{|\vec{q}|^2}-
\frac{(\vec{p} \cdot \vec{q})^2}{|\vec{q}|^4}\right] \delta^{(3)} (\vec{q})\,.
\label{VEVuno}
\end{eqnarray}
and consider the Taylor expansion of $\varphi^*_{|\vec{p}-\vec{q}|}$ around $\vec{q}=0$ 
(we keep only the terms that can give a non zero contribution to the final result)
\be
\varphi^*_{|\vec{p}-\vec{q}|}=\varphi^*_p-\frac{\vec{q} \cdot \vec{p}}{p} \frac{\partial}{\partial p}
\varphi^*_p+\frac{1}{2}\left\{\left[-\frac{(\vec{q} \cdot \vec{p})^2}{p^3}+
\frac{q^2}{p}\right] \frac{\partial}{\partial p} \varphi^*_p+\frac{(\vec{q} \cdot \vec{p})^2}{p^2} 
\frac{\partial^2}{\partial p^2} \varphi^*_p
\right\} \,.
\label{ExpVarphi}
\ee
Substituting such expansion in (\ref{VEVuno}) one gets
\begin{eqnarray}
& & \left\langle \hat{O}^{ij} \partial_i \vf\, \partial_j \vf \right\rangle= \int \frac{d^3 {\bf p}}{(2 \pi)^3} d^3 {\bf q}
 \, e^{i \vec{q} \cdot \vec{x}} \delta^{(3)} (\vec{q})\,
\left\{ 
|\varphi_p|^2 \left[\frac{1}{3}\frac{2 (\vec{p} \cdot \vec{q})+|\vec{p}|^2}{|\vec{q}|^2}-
\frac{(\vec{p} \cdot \vec{q})^2}{|\vec{q}|^4}\right] \right. \nonumber \\
& & \left. \,\,\,\,\,\,\,\,\,\,\,\,\,\,\,\,\,\,\,\,\,\,\,\,\,\,\,\,
+\varphi_p \frac{\partial}{\partial p} \varphi^*_p 
\left[\frac{p}{6}+\frac{1}{3} \left(\frac{1}{p}-\frac{p}{q^2}\right) (\vec{p} \cdot \vec{q})-\frac{4}{3} 
\frac{(\vec{p} \cdot \vec{q})^2}{p q^2}+
\left(\frac{1}{p q^4}-\frac{1}{3 p^3 q^2}\right)
(\vec{p} \cdot \vec{q})^3+\frac{1}{2} \frac{(\vec{p} \cdot \vec{q})^4}{p^3 q^4}
\right]\right. \nonumber \\
& & \left. \,\,\,\,\,\,\,\,\,\,\,\,\,\,\,\,\,\,\,\,\,\,\,\,\,\,\,\,
 +\varphi_p \frac{\partial^2}{\partial p^2} \varphi^*_p 
\left[\frac{1}{6}\frac{(\vec{p} \cdot \vec{q})^2}{q^2}
+\frac{1}{3}\frac{(\vec{p} \cdot \vec{q})^3}{p^2 q^2}
-\frac{1}{2}\frac{(\vec{p} \cdot \vec{q})^4}{p^2 q^4}\right]
\right\}\,,
\label{VEVdue}
\end{eqnarray} 
which is, in fact, a real quantity.

Let us decompose in polar coordinates the $\vec{p}$ measure inside (\ref{VEVdue}):
\begin{eqnarray}
\left\langle \hat{O}^{ij} \partial_i \vf\, \partial_j \vf \right\rangle&=&  
\int \frac{d^3 {\bf p}}{(2 \pi)^3} d^3 {\bf q}
 \, e^{i \vec{q} \cdot \vec{x}} \delta^{(3)} (\vec{q})\,
\left\{.......\right\}
\nonumber \\
& &\!\!\!\!\!\!\!\!\!\!\!\!\!\!\!\!\!\!\!\!\!\!\!\!\!\!\!\!\!\!\!\!\!=
\int \frac{d^3 {\bf q}}{(2 \pi)^3} \delta^{(3)} (\vec{q})  e^{i \vec{q} \cdot
  \vec{x}}
\int_0^{+\infty} dp \, p^2 \int_0^{2 \pi}d \phi
\int_0^\pi d \theta \sin \theta  \left\{.......\right\}\,,
\label{VEVtre}
\end{eqnarray} 
where $\theta$ is the angle between $\vec{q}$ and $\vec{p}$.
Performing the integration with respect to $\theta$, it is easy 
to see that the terms that can give an infrared divergent contribution, as a
consequence of the delta function, cancel.

In particular, looking at the integrand in \eqref{VEVdue}, one finds that after
the $\theta$ integration the coefficient of $|\varphi_p|^2$ is not only
finite, but simply zero.

After the $\theta$ and $\phi$ integral are done, one is left with
\begin{eqnarray}
\left\langle \hat{O}^{ij} \partial_i \vf\, \partial_j \vf \right\rangle&=&  
\int \frac{d^3 {\bf q}}{(2 \pi)^2} \delta^{(3)} (\vec{q})  e^{i \vec{q} \cdot
  \vec{x}}
\int_0^{+\infty} \!\!\!\!dp \, p^2  \varphi_p \left[-\frac{16}{45}
 \, p\frac{\partial}{\partial p} \varphi^*_p -\frac{4}{45} \,p^2\frac{\partial^2}{\partial p^2} \varphi^*_p \right]\,,
\label{VEVquattro}
\end{eqnarray} 
where the $\vec{q}$ integration can be trivially performed.

Let us note as expression (\ref{VEVquattro}), once renormalization is considered \cite{FMVV_II,starall}, become finite with contribution of modes essentially with momentum support in the range
$c_1 H_0<p<c_1 a H$, with $c_1$ a constant such that $c_1\ll 1$. 
As a consequence we can consider (see \cite{BPT}) the following expression for the mode functions (in the slow-roll 
limit and on large scales)
\be
\varphi_p = -\frac{i}{H}\frac{H(t_p)^2}{\sqrt{2 p^3}}
\ee
with
\be
H(t_p)\simeq H_0 \left(1-2 \epsilon_0 \ln \frac{p}{H_0}\right)^{1/2}
\ee
which is the Hubble parameter when the fluctuations crosses the horizon and $\epsilon_0$ is the value 
of $\epsilon$ at the beginning of inflation.
In this case we have that
\be
\frac{\partial}{\partial p} \varphi^*_p=-\frac{3}{2} \frac{\varphi^*_p}{p}+2 \epsilon_0 
\frac{\varphi^*_p}{p} \left(\frac{H_0}{H(t_p)}\right)^2
\ee
and the first term of Eq.(\ref{VEVquattro}) can be written as 
\be
\frac{1}{4 \pi^2} \int_{c_1 H_0}^{c_1 a H} \!\!\!\!dp \, p^2  \left(-\frac{16}{45}\right)
\left[-\frac{3}{2}|\varphi_p|^2+2 \epsilon_0 \frac{H_0}{H(t_p)^2}  |\varphi_p|^2\right]
\ee 
following the result of \cite{starall} we obtain for this last expression the following result
\be 
\frac{1}{30 \pi^2} \frac{1}{m^2} \left(\frac{H_0^6}{H^2}-H^4\right)+ 
\frac{1}{15 \pi^2} \frac{1}{m^2} \left(-\frac{\dot{H}}{H^2}\right)\left(H_0^4-H^4\right)
\ee
and the second term is subleading.
Similar results can be obtained considering the term with the second derivative of $\varphi_p^*$ and, 
as a consequence, to obtain a reasonable estimate we can assume that the mode functions are roughly
$\varphi_p \sim 1/p^{3/2}$. We then obtain
a leading behavior $\varphi_p \, p\frac{\partial}{\partial p} \varphi^*_p\simeq -\frac{3}{2}
|\varphi_p|^2$
and $\varphi_p \,p^2\frac{\partial^2}{\partial p^2} \varphi^*_p \simeq
\frac{15}{4}|\varphi_p|^2 $ which leads to the following renormalized result

\begin{eqnarray}
\left\langle \hat{O}^{ij} \partial_i \vf\, \partial_j \vf \right\rangle&=&  
\frac{1}{(2 \pi)^2}
\int_{c_1 H_0}^{c_1 a H} \!\!\!\!dp \, p^2  \left[\frac{8}{15} |\varphi_p|^2
  -\frac{1}{3}|\varphi_p|^2 \right]
= \frac{1}{10} \langle \vf^2\rangle \,.
\label{VEVcinque}
\end{eqnarray} 


\begin{thebibliography}{999}
\newcommand{\bb}{\bibitem}

\bibitem{ReviewsBE} 
T.~Buchert,
  arXiv:1103.2016 [gr-qc];
G.~F.~R.~Ellis,
  arXiv:1103.2335 [astro-ph.CO].

\bibitem{Uc}
  W.~Unruh,
  arXiv:astro-ph/9802323.

\bibitem{GE}
G. F. R. Ellis, in ``{\em 10th International Conference on 
General Relativity and Gravitation}'', 
ed. by B. Bertotti, De Felice and A. Pascolini 
(Reidel, Dordrecht, 1984), p. 215.

\bibitem{Buchert} T. Buchert, Gen. Rel. Grav. {\bf 32}, 105 (2000).

\bibitem{FMVVprl}
  F.~Finelli, G.~Marozzi, G.~P.~Vacca and G.~Venturi,
  Phys.\ Rev.\ Lett.\  {\bf 106}, 121304 (2011).

\bibitem{ABM}
  V.~F.~Mukhanov, L.~R.~W.~Abramo and R.~H.~Brandenberger,
  Phys.\ Rev.\ Lett.\  {\bf 78}, 1624 (1997);
  L.~R.~W.~Abramo, R.~H.~Brandenberger and V.~F.~Mukhanov,
  Phys.\ Rev.\  D {\bf 56} (1997) 3248.

\bibitem{All}
L.~R.~W.~Abramo and R.~P.~Woodard,
  Phys.\ Rev.\  D {\bf 60} (1999) 044010;
  Phys.\ Rev.\  D {\bf 65} (2002) 063515;
  Phys.\ Rev.\  D {\bf 65}, 043507 (2002).
  G.~Geshnizjani and R.~Brandenberger,
  Phys.\ Rev.\  D {\bf 66}, 123507 (2002);
B.~Losic and W.~G.~Unruh,
  Phys.\ Rev.\  D {\bf 72}, 123510 (2005).

\bibitem{FMVV_II}
F.~Finelli, G.~Marozzi, G.~P.~Vacca, and G.~Venturi,
Phys.\ Rev.\ D {\bf 69}, 123508 (2004).

\bibitem{GMV1}
M. Gasperini, G. Marozzi, and G. Veneziano, JCAP {\bf 03}, 011 (2009).

\bibitem{GMV2}
M. Gasperini, G. Marozzi, and G. Veneziano, JCAP {\bf 02}, 009 (2010).


\bibitem{MetAll}
  M.~Bruni, S.~Matarrese, S.~Mollerach and S.~Sonego,
  Class.\ Quant.\ Grav.\  {\bf 14}, 2585 (1997).

\bibitem{Mar}
 G.~Marozzi,
 JCAP {\bf 01}, 012 (2011).
 
 
 \bibitem{GMNV}
  M.~Gasperini, G.~Marozzi, F.~Nugier and G.~Veneziano,
  JCAP {\bf 07}, 008 (2011).

\bb{BE} T.  Buchert and J. Ehlers, Astron. Astrophys. {\bf 320}, 1 (1997).

\bibitem{Larena} J.~Larena,
  Phys.\ Rev.\  D {\bf 79} (2009) 084006.

\bibitem{EvE}
  G.~F.~R.~Ellis, H.~van Elst,
  NATO Adv.\ Study Inst.\ Ser.\ C.\ Math.\ Phys.\ Sci.\  {\bf 541}, 1-116 (1999).
  
\bibitem{PeterUzan}
P. Peter, J.-P. Uzan, "Primordial Cosmology", Oxford University Press (2009).

\bibitem{Mukhanov}
V. F. Mukhanov, JETP Lett. {\bf 41}, 493 (1985);
Sov.\ Phys.\ JETP {\bf 68} 1297 (1988).


\bibitem{starall2}
F.~Finelli, G.~Marozzi, A. A. Starobinsky, G.~P.~Vacca and G.~Venturi,
Phys.\ Rev.\  D {\bf 82}, 064020 (2010).

\bibitem{starall}
F.~Finelli, G.~Marozzi, A. A. Starobinsky, G.~P.~Vacca and G.~Venturi,
Phys. Rev. D {\bf 79}, 044007 (2009).

\bibitem{BPT} 
F.~Finelli, G.~Marozzi, G.~P.~Vacca and G.~Venturi,
Phys. Rev. {\bf D 74}, 083522 (2006);





\end{thebibliography}
\end{document}